\documentstyle[epsfig]{aipproc}

\def\agt{\mathrel{\raise.3ex\hbox{$>$}\mkern-14mu\lower0.6ex\hbox{$\sim$}}}
\def\alt{\mathrel{\raise.3ex\hbox{$<$}\mkern-14mu\lower0.6ex\hbox{$\sim$}}}
\newcommand{\beq}{\begin{equation}}
\newcommand{\eeq}{\end{equation}}
\newcommand{\beqn}{\begin{eqnarray}}
\newcommand{\eeqn}{\end{eqnarray}}
\newcommand{\varep}{\varepsilon}
\newcommand{\pa}{\partial}
\def\agt{\mathrel{\raise.3ex\hbox{$>$}\mkern-14mu\lower0.6ex\hbox{$\sim$}}}

\begin{document}

\title{\vspace{-1cm}
Binary neutron star mergers in fully general 
relativistic simulations} 

\author{\vspace{-6mm} Masaru Shibata$^*$ and K\=oji Ury\=u$^{\dagger}$}
\vspace{-6mm}
\address{$^*$Graduate School of Arts and Sciences, 
University of Tokyo, Komaba, Meguro 153-8902, Japan \\
$^{\dagger}$Department of Physics, University of Wisconsin-Milwaukee, 
P.O. Box 413, Milwaukee, WI53201, USA}

%\lefthead{LEFT head}
%\righthead{RIGHT head}
\maketitle

\vspace{-9mm}
\begin{abstract}
We perform 3D numerical simulations for merger of equal mass 
binary neutron stars in full general relativity 
preparing irrotational 
binary neutron stars in a quasiequilibrium state as initial conditions. 
Simulations have been carried out for a wide range of stiffness of 
equations of state and compactness of neutron stars, 
paying particular attention to the final products and gravitational waves. 
We take a fixed uniform grid in Cartesian coordinates with typical grid size 
$(293,293,147)$ in $(x,y,z)$ assuming 
a plane symmetry with respect to the equatorial plane. 
A result of one new large-scale simulation performed with grid size 
$(505,505,253)$ is also presented. 

We find that the final product depends sensitively on the initial 
compactness of the neutron stars: 
In a merger between sufficiently compact neutron stars, a black hole is 
formed in a dynamical timescale.  As the compactness is decreased, 
the formation timescale becomes longer and longer.  
For less compact cases, a differentially rotating 
massive neutron star is formed instead of a black hole. 
In the case of black hole formation, 
the disk mass around the black hole appears to be smaller 
than $1\%$ of the total rest mass. It is also indicated that 
waveforms of high-frequency gravitational waves after merger depend 
strongly on the compactness of neutron stars. 
\end{abstract}

%%%\vspace{-4mm}
\section{Introduction}

Interest in the merger phase of binary neutron stars 
has been stimulated by the prospect 
of future observation of extragalactic, close binary 
neutron stars by gravitational wave detectors \cite{KIP}. 
A statistical study indicates that mergers of binary neutron stars 
may occur at a few events 
per year within a distance of a few hundred Mpc \cite{phinney}. 
Since the amplitude of gravitational waves from a binary 
of mass $\sim 3M_{\odot}$ (where $M_{\odot}$ denotes the solar mass) 
at a distance of $\sim 100$Mpc can be $\sim 10^{-22}-10^{-21}$, the 
merger of binaries is a promising source for gravitational wave detectors. 
Although the frequency of gravitational waves in the 
merging regime will be larger than 1kHz (see Sec. IV) and hence 
lies beyond the upper end of an accessible frequency 
range of laser interferometers such as LIGO 
for a typical event at a distance $\sim 100$Mpc,  
it may be observed using specially designed 
narrow band interferometers or resonant-mass detectors \cite{KIP}. 
Future observations, although they may not be 
done in near future, will provide valuable information about 
merger mechanisms and final products of binary neutron stars. 

Interest has also been stimulated by a hypothesis about 
the central engine of $\gamma$-ray bursts (GRBs) \cite{piran}. Recently, 
some of GRBs have been found to be of cosmological 
origin \cite{grb}. In cosmological GRBs, 
the central sources must supply 
a large amount of the energy $\geq 10^{50}$ ergs in a 
very short timescale of order from a millisecond to a second. 
It has been suggested that 
the merger of binary neutron stars is a likely candidate 
for the powerful central source \cite{piran}. 
In the typical hypothetical scenario, 
the final product should be a system composed of a rotating 
black hole surrounded by a massive disk of mass $>0.1M_{\odot}$, 
which could supply a large amount of energy by neutrino processes 
or by extracting the rotational energy of the black hole \cite{piran}. 

To investigate merger of binary neutron stars theoretically, 
numerical simulation appears to be unique promising approach. 
Considerable effort has been made for this in the 
framework of Newtonian and post-Newtonian approximation 
(see, e.g., \cite{NOS}). 
Although these simulations have clarified a wide variety 
of physical features which are important during the 
merger of binary neutron stars, a fully general relativistic 
(GR) treatment is obviously necessary for determining the final product 
and associated gravitational waves because GR effects are crucial. 

Effort has been made for constructing a reliable 
code for 3D hydrodynamic simulation 
in full general relativity (see, e.g., \cite{ON,waimo,gr3d}). 
Recently, the authors have succeeded in constructing a 
numerical code in which stable and fairly accurate simulations are 
feasible \cite{gr3d}. Applying the newly developed code, the authors 
and collaborators have been performing fully GR 
simulations for a wide variety of astrophysical problems 
\cite{gr3d,rot1,rot2,bina,rot3}. 
In this paper, we present some of our recent results 
with regard to the merger of binary neutron stars.

The paper is organized as follows. In Sec. II, we summarize 
our formulation in 3D numerical relativity. 
In Sec. III, we briefly describe how to give initial conditions. 
In Sec. IV, numerical results with regard to 
merger of binary neutron stars are presented, paying particular attention 
to the final product and gravitational waveforms. 
Sec. V is devoted to a summary. 
Throughout this paper, we adopt the units $G=c=1$ where $G$ and $c$
denote the gravitational constant and speed of light, respectively.  
We use Cartesian coordinates $x^k=(x, y, z)$ 
as the spatial coordinates; $t$ denotes coordinate time. 
In the following, 
BH and NS denote black hole and neutron star, respectively. 

\section{Summary of the Formulation}

We perform hydrodynamic simulations in full general relativity using 
(3+1) formalism. We use the same formulation and gauge conditions 
as in our previous works \cite{gr3d,gw3p2}, 
to which the reader may refer for details and basic equations. 
The fundamental variables used in this paper are 
$\rho$: rest mass density, $\varep$: specific internal energy, 
$P$:  pressure, $u^{\mu}$: four velocity, $v^{k}=u^k/u^t$, 
$\Omega =v^{\varphi}$, $\alpha$: lapse function, 
$\beta^k$: shift vector, $\gamma_{ij}$: three metric, 
$\gamma=e^{12\phi}={\rm det}(\gamma_{ij})$, 
$\tilde \gamma_{ij}=e^{-4\phi}\gamma_{ij}$, 
$K_{ij}$: extrinsic curvature. 
Definitions of several quantities are also found in Table 1. 

Geometric variables, $\phi$, $\tilde \gamma_{ij}$, 
the trace of the extrinsic curvature $K_k^{~k}\equiv K_{ij}\gamma^{ij}$, 
a tracefree part of the extrinsic curvature 
$\tilde A_{ij}\equiv e^{-4\phi}(K_{ij}-\gamma_{ij}K_k^{~k}/3)$, 
as well as three auxiliary functions $F_i \equiv
\pa_j \tilde \gamma_{ij}$, where $\pa_j$ is the partial derivative, 
are evolved with an unconstrained evolution code in a modified form of
the ADM formalism \cite{SN}.  GR hydrodynamic
equations are evolved using a van Leer scheme for the advection terms 
\cite{Hydro} adding artificial viscous terms \cite{gr3d}. 
Violations of the
Hamiltonian constraint and conservation of mass and angular momentum
are monitored to check numerical accuracy.  
Reliability of the 
code has been checked by several test calculations, including
spherical collapse of dust, stability of spherical neutron stars, and
the stable evolutions of rigidly and rapidly rotating neutron stars
which have been described in \cite{gr3d}.  

We adopt a $\Gamma$-law equation of state $~P=(\Gamma-1)\rho \varep~$ 
where $\Gamma$ is the adiabatic constant.  For isentropic configurations 
the equation of state can be rewritten in the polytropic
form $~P = \kappa \rho^{\Gamma}~$ and $~\Gamma = 1 + 1/n~$ 
where $\kappa$ is the polytropic constant and $n$ the polytropic
index.  This is the form that we use for constructing quasiequilibrium 
states as initial conditions.
We adopt $n = 2/3$, 4/5, 1, and $5/4$ as a reasonable qualitative
approximation to a moderately stiff, nuclear equation of state.

Instead of $\rho$ and $\varep$ we numerically evolve the densities
$\rho_* \equiv \rho \alpha u^0 e^{6\phi}$ and $e_* \equiv
(\rho\varepsilon)^{1/\Gamma}\alpha u^0 e^{6\phi}$ as the hydrodynamic
variables \cite{gr3d}. In our numerical method, 
the total rest mass of the system 
\beq 
M_*=\int d^3 x \rho_*.  
\eeq 
is automatically conserved. 

The time slicing and spatial gauge conditions we use in this paper for the 
lapse and shift are the
same as those adopted in our series of papers 
\cite{gr3d,rot1,rot2,bina,rot3,gw3p2}; i.e.~we impose an ``approximate''
maximal slice condition ($K_k^{~k} \simeq 0$) and an ``approximate''
minimum distortion gauge condition ($\tilde D_i (\pa_t \tilde
\gamma^{ij}) \simeq 0$ where $\tilde D_i$ is the covariant derivative
with respect to $\tilde \gamma_{ij}$; see \cite{gw3p2}).

\section{Initial condition}

Even just before merger, binary neutron stars are 
considered to be in a quasiequilibrium state because 
the timescale of gravitational radiation reaction 
$\sim 5/\{64\Omega(M_g\Omega)^{5/3}\}$ \cite{ST}, 
where $M_g$ and $\Omega$ denote the total mass of the system and 
the orbital angular velocity of the binary neutron stars, 
is several times longer than the orbital period (cf. Table 1). Thus, 
for a realistic simulation of the merger, 
we should prepare a quasiequilibrium state as the initial condition. 

Since the viscous timescale in the neutron star is much longer than 
the evolution timescale associated with gravitational radiation, 
the vorticity of the system conserves in the late inspiraling phase of 
binary neutron stars \cite{CBC}. Furthermore, 
the orbital period just before the merger 
is about 2msec which is much shorter than the spin period of 
typical neutron stars, implying that even if the spin of 
neutron star would exist at a distant orbit and 
conserve throughout the subsequent evolution, 
it could be negligible at the merger phase in most cases. 
Thus, it is quite reasonable to assume 
that the velocity field of neutron stars just before the merger 
is irrotational.

To prepare quasiequilibrium states, we 
assume the existence of a helicoidal Killing vector in the form, 
$\ell^{\mu}=(1, -y\Omega, x\Omega, 0)$. For irrotational fluid, 
the hydrostatic equation is integrated to give a 
Bernoulli type equation in the presence of 
$\ell^{\mu}$\cite{irre}, resulting in a great simplification 
for handling the hydrodynamic equations. 

%%Since emission of gravitational waves violates the 
%%helicoidal symmetry, this assumption does not strictly hold. 
%%However, the emission timescale of gravitational waves 
%%is several times 
%%longer than the orbital period even just before the merger 
%%({\it cf}. Table I) so that this assumption is acceptable 
%%for computing an approximate quasiequilibrium state. 

Currently, we restrict our attention to 
initial conditions satisfying $\tilde \gamma_{ij}=\delta_{ij}$, 
$\pa_t \tilde\gamma_{ij}=0$ and $K_k^{~k}=0$. 
The initial conditions for geometric 
variables are obtained by solving the Hamiltonian and 
momentum constraint equations, and equations for 
gauge conditions. 
In this case, the basic equations reduce to 
two scalar elliptic type equations for $\alpha$ and $\psi(=e^{\phi})$ 
and one vector elliptic type equation for $\beta^k$ \cite{UE}. 

The coupled equations of Bernoulli type equation and elliptic 
type equations for metric are solved using the method 
developed by Ury\=u and Eriguchi \cite{UE}. 
Several test results and scientific results are found 
in \cite{UE} and \cite{USE}.

\section{Results}

%%%%%%%%%%%%%%%%%%%%%%%%%%%%%%%%%%%%%%%%%%%%%%%%%%%%%%%%%%%%%%%
\begin{table}[t]
{\footnotesize
\begin{center}
\begin{tabular}{|c|c|c|c|c|c|c|c|c|c|c|c|} \hline
Model &
\hspace{0mm} $C$ \hspace{0mm} &
\hspace{1mm} $\bar \rho_{\rm max} $ \hspace{1mm} &
\hspace{1mm} $\bar M_*$ \hspace{1mm} & 
\hspace{1mm} $\bar M_{g}$\hspace{1mm} 
& \hspace{1mm} $q$ \hspace{1mm} & 
\hspace{1mm} $X$ \hspace{1mm}
& \hspace{1mm} $R_{\tau}$ \hspace{1mm}  & 
$C_{\rm mass}$ & $L/M_g$ & Products 
\\ \hline
(A)&0.12 & 0.139  & 0.186 & 0.173 & 1.03 & 0.090 & 5.1 &0.58 & 37.5 
& NS \\ \hline
(B)&0.14 & 0.169  & 0.216 & 0.198 & 0.98 & 0.106 & 3.4 &0.67 & 31.1
& marginal  \\ \hline
(C)&0.16 & 0.202  & 0.244 & 0.220 & 0.93 & 0.124 & 2.3 &0.75 & 26.3 
& BH  \\ \hline
\end{tabular}
\caption{\footnotesize 
A list of quantities for 
initial conditions of irrotational binary neutron stars with $\Gamma=2.25$. 
The compactness of each star in isolation $C\equiv (M/R)_{\infty}$, 
the maximum density $\bar \rho_{\rm max}\equiv \kappa^n \rho_{\rm max}$, 
total rest mass $\bar M_* \equiv \kappa^{-n/2} M_*$, 
gravitational mass at $t=0$ $\bar M_{g} \equiv \kappa^{-n/2} M_g$, 
$q\equiv J/M_{g}^2$ where $J$ is the total angular momentum, 
compactness of orbit 
$X \equiv (M_{g}\Omega)^{2/3}(\sim M_{g}/a$ where $a$ is orbital separation), 
approximate ratio of the emission timescale of gravitational waves to 
the orbital period $R_{\tau}\equiv 5(M_{g}\Omega)^{-5/3}/128\pi$, 
ratio of the rest mass of each star to 
the maximum allowed mass for a spherical star 
$C_{\rm mass}\equiv M_*/2M_{*~\rm max}^{\rm sph}$, 
location of outer boundaries $L$ along three axes in units of $M_g$ 
(i.e., $L/M_g$) for simulations with $(293,293,147)$ 
grid size, and final products are shown. 
All quantities are normalized by $\kappa$ to be 
non-dimensional: We can rescale by appropriately choosing $\kappa$. 
Here, $M_{*~\rm max}^{\rm sph}$ denotes the maximum allowed mass 
of a spherical star ($\kappa^{-n/2}M_{*~\rm max}^{\rm sph} \simeq 0.162$ 
at $\bar \rho_{\rm max} \simeq 0.52)$. }
\end{center}
}
\vspace{-6mm}
\end{table}
%\end{aiptable}
%%%%%%%%%%%%%%%%%%%%%%%%%%%%%%%%%%%%%%%%%%%%%%%%%%%

We have performed simulations for $\Gamma=1.8$, 2.0, 2.25 
and 2.5. The results for $\Gamma=2$ 
have been already presented in \cite{bina}. 
In this manuscript, we show results for $\Gamma=2.25$ which 
have recently obtained. 
In Table 1, we list several quantities which characterize the 
quasiequilibrium state of irrotational binary neutron stars 
used as initial conditions. 
All the quantities are 
scaled with respect to $\kappa$ (in units of $c=G=1$) 
to be non-dimensional. 
We choose binaries at the innermost 
orbits for which the Lagrange points appear at the inner edge of 
neutron stars \cite{USE}. To induce merger, 
we reduce the angular momentum $J$ by $\sim 2-3\%$ (see discussion below). 
The gravitational mass $M_g$ and non-dimensional angular momentum 
parameter $q\equiv J/M_g^2$ listed in Table 1 are calculated from 
initial data sets which are recomputed by solving 
constraint equations after the reducing of $J$. 

We have performed simulations using a fixed uniform grid 
assuming reflection symmetry with respect to the equatorial plane. 
The simulations were mainly performed using FACOM VPP300/16R. 
In this case, the typical grid size was $(293,293,147)$ in $(x,y,z)$. 
One large-scale 
simulation was recently performed using FACOM VPP5000 with 
$(505,505,253)$ grid size to enlarge 
the computational region. In both cases, 
the grid spacing is identical and determined 
from the condition that major diameter of each star is covered 
with $\sim 33$ grid points initially. 
The computational time for one simulation was typically $\sim 100$CPU hours 
for $\sim 10^4$ timesteps. Test simulations were performed 
with $(193,193,97)$ grid size on FACOM VX/4R 
to check the convergence of numerical results. 

The wavelength of gravitational waves at $t=0$ is computed as 
\beq
\lambda_{t=0} \equiv {\pi \over \Omega} \simeq 100M_g 
\biggl({X \over 0.1}\biggr)^{-3/2},
\eeq
where $X\equiv (M_g\Omega)^{2/3}$ 
denotes a compactness of orbits (cf. Table 1). 
To accurately extract the waveform near outer boundaries, 
$\lambda_{t=0}$ should be shorter than the size of 
the computational region along three axes $L$. 
However, as found in Table 1, this condition is not satisfied 
since taking 
such a large computational region is a very difficult task in the 
present restricted computational resources: 
$\lambda_{t=0}$ is typically $\sim L/3$ with $(293,293,147)$ 
grid size. Even with $(505,505,253)$ grid size, 
$\lambda_{t=0}$ is $\sim 2L/3$. This implies 
that gravitational waves in the early phase cannot be accurately computed. 
However, the wavelength of quasi-periodic 
waves of the merged object excited during merger (see Fig. 6) 
is much shorter than the wavelength of binaries in 
quasiequilibrium and $L$. Therefore, 
the waveforms in the late phase to which we here pay main attention 
can be computed fairly accurately. 

As found in \cite{USE}, 
orbits of all irrotational binaries with $\Gamma \alt 2.25$ 
are stable and the merger in reality should be triggered by 
radiation reaction of gravitational waves. To 
take into account the gravitational radiation reaction approximately, 
we use the following method. Using the quadrupole formula, we 
can estimate the angular momentum loss in one orbital period 
$\Delta J (>0)$ as $4\pi M_g X^2/5$ \cite{ST} 
and hence can write the ratio of $\Delta J$ to the total angular momentum 
as
\beq
{\Delta J \over J}
= {4\pi \over 5q}X^2=0.025\biggl({1 \over q}\biggr)
\biggl({X \over 0.1}\biggr)^2. 
\eeq
Thus, we initially reduce the angular momentum by $\sim 2-3\%$.

\begin{figure}[t]
\begin{center}
\epsfxsize=1.8in
\leavevmode
\epsffile{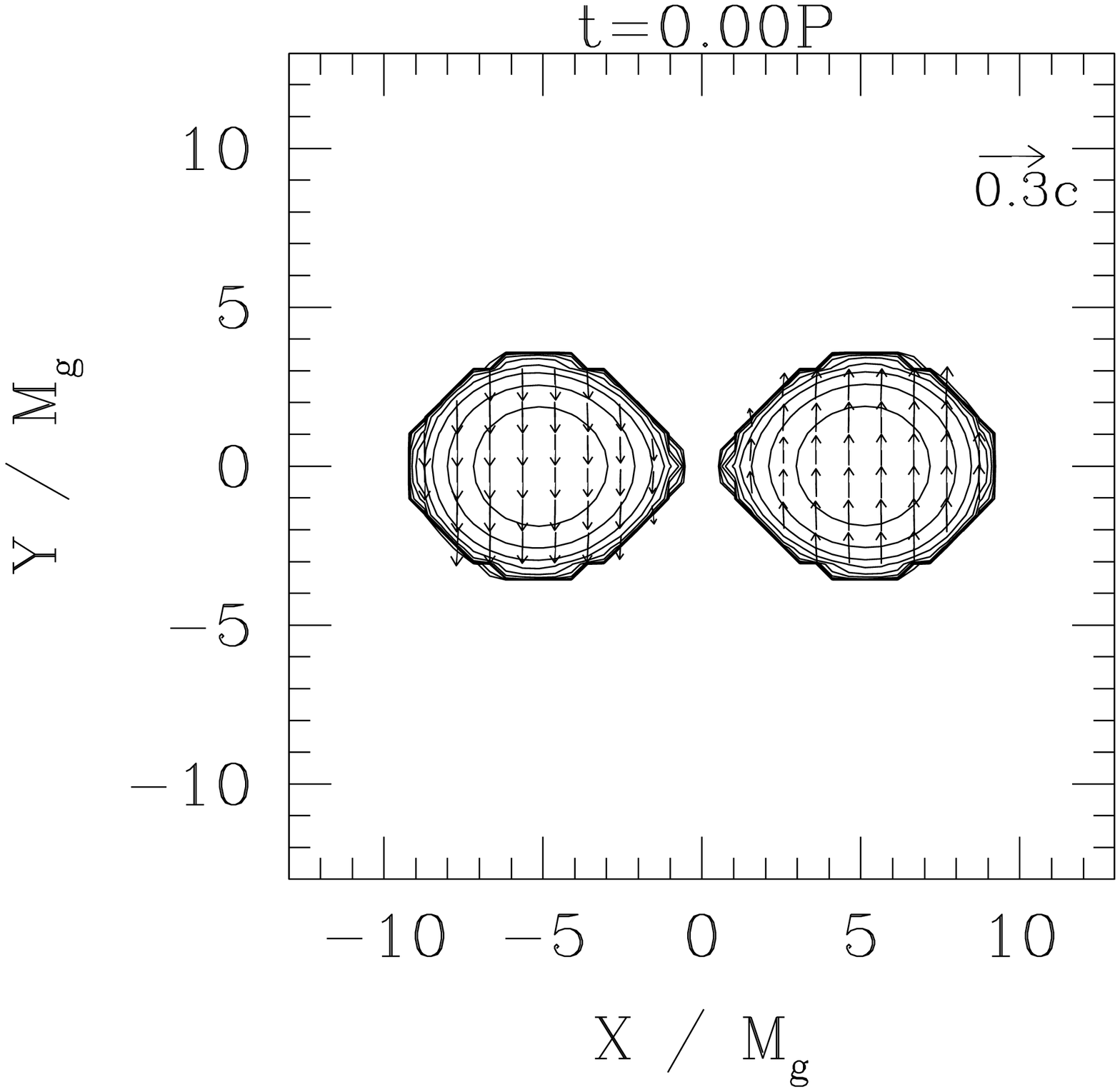}
\epsfxsize=1.8in
\leavevmode
\epsffile{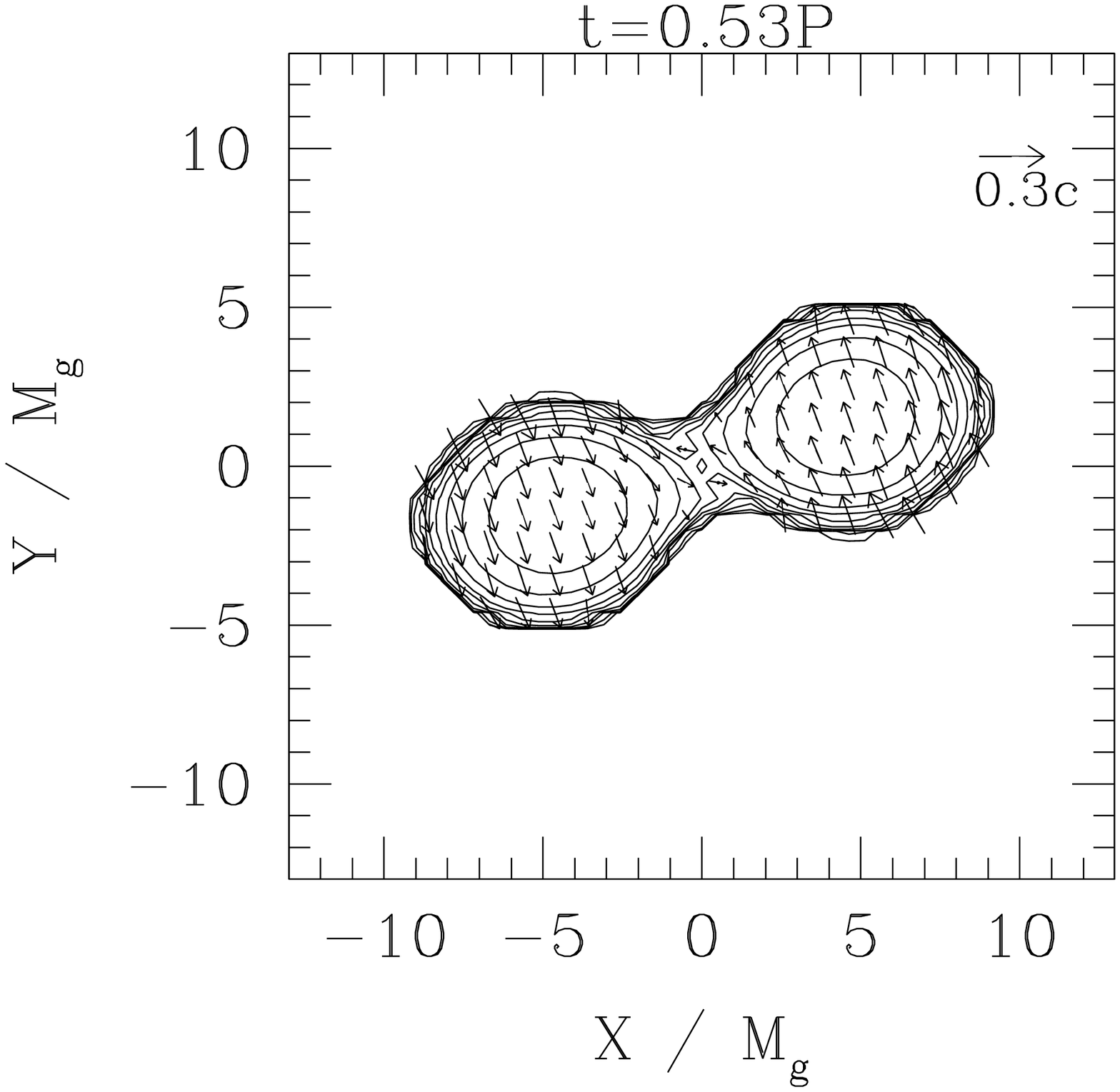} 
\epsfxsize=1.8in
\leavevmode
\epsffile{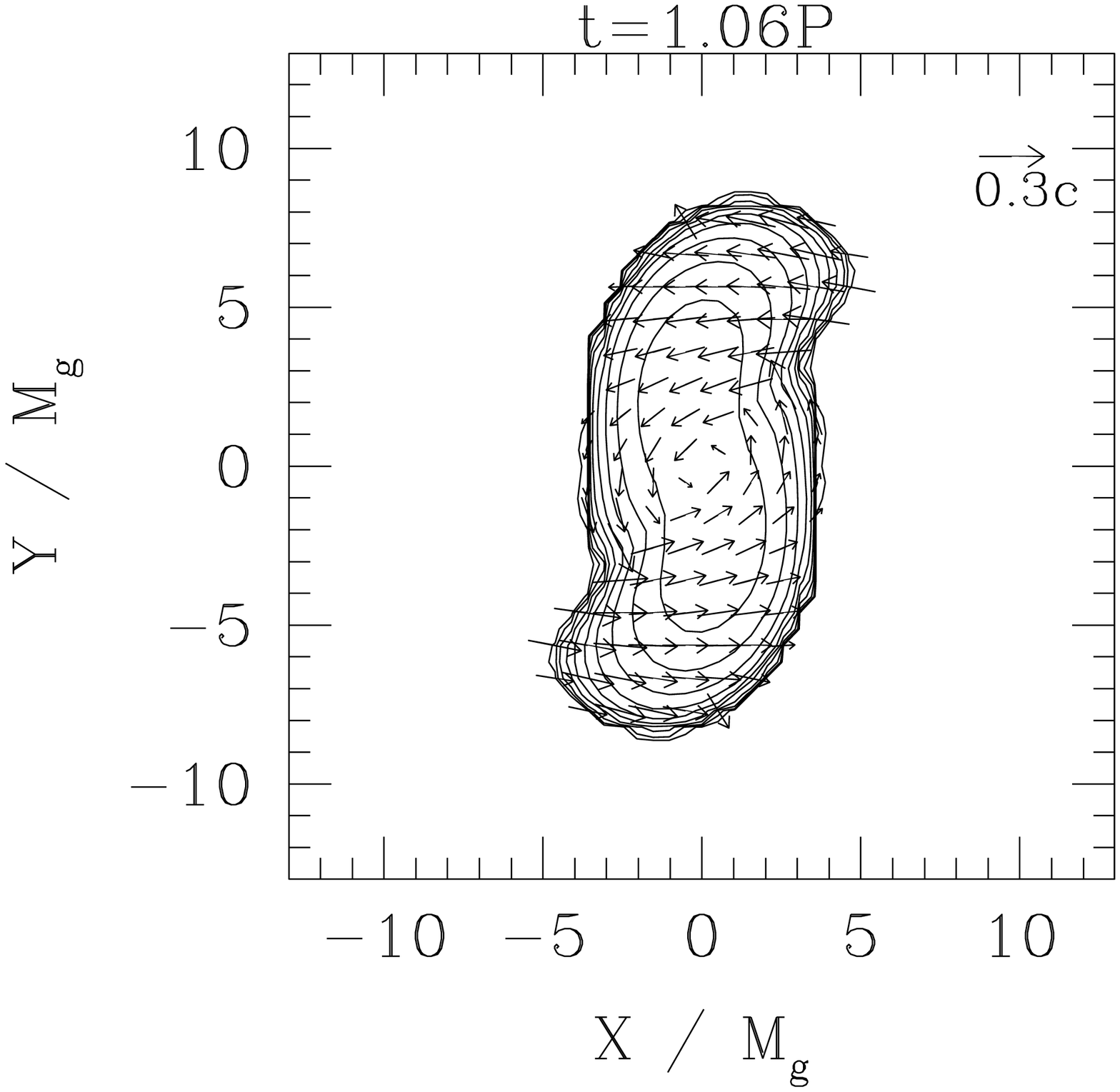} \\
\epsfxsize=1.8in
\leavevmode
\epsffile{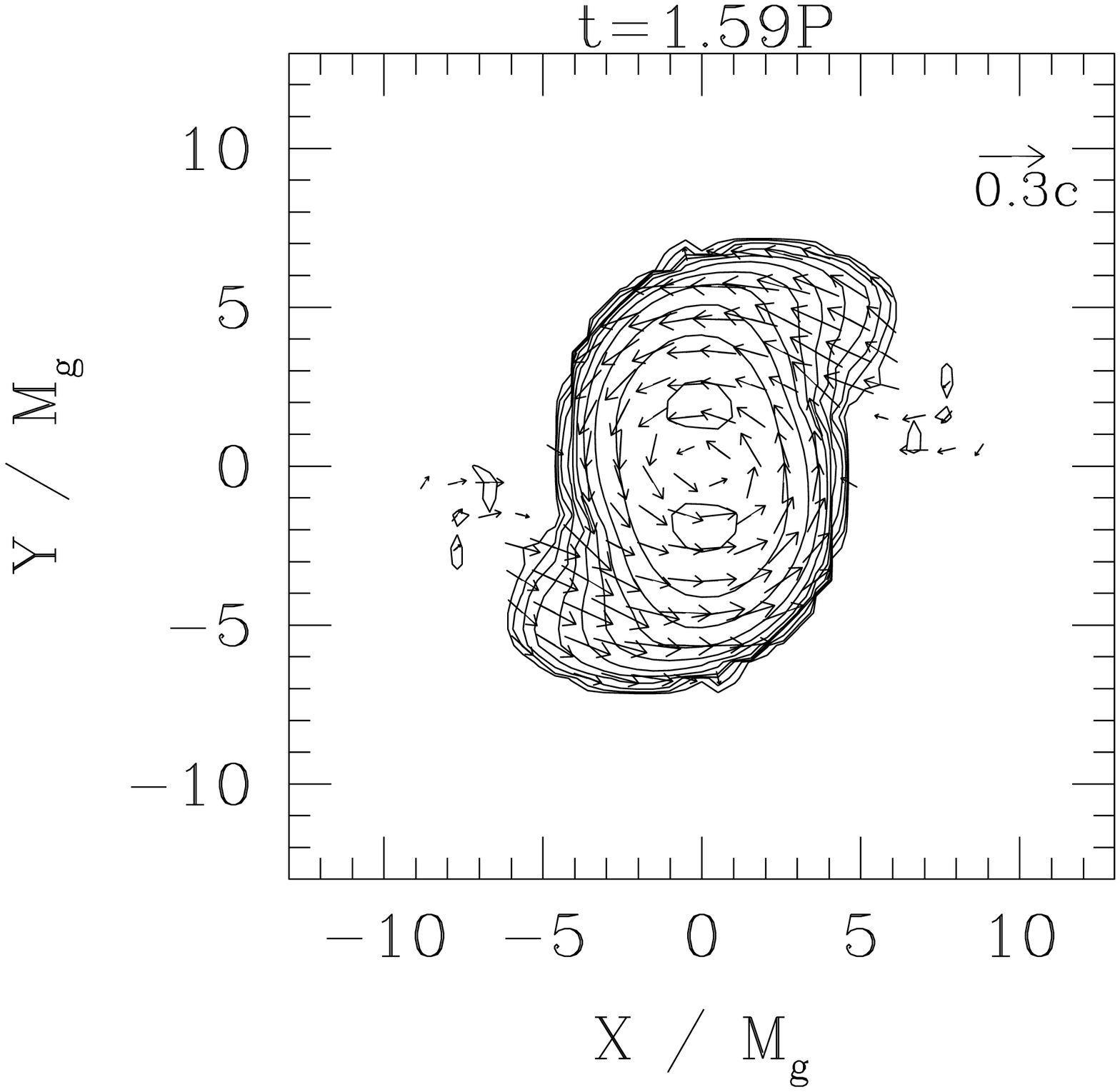} 
\epsfxsize=1.8in
\leavevmode
\epsffile{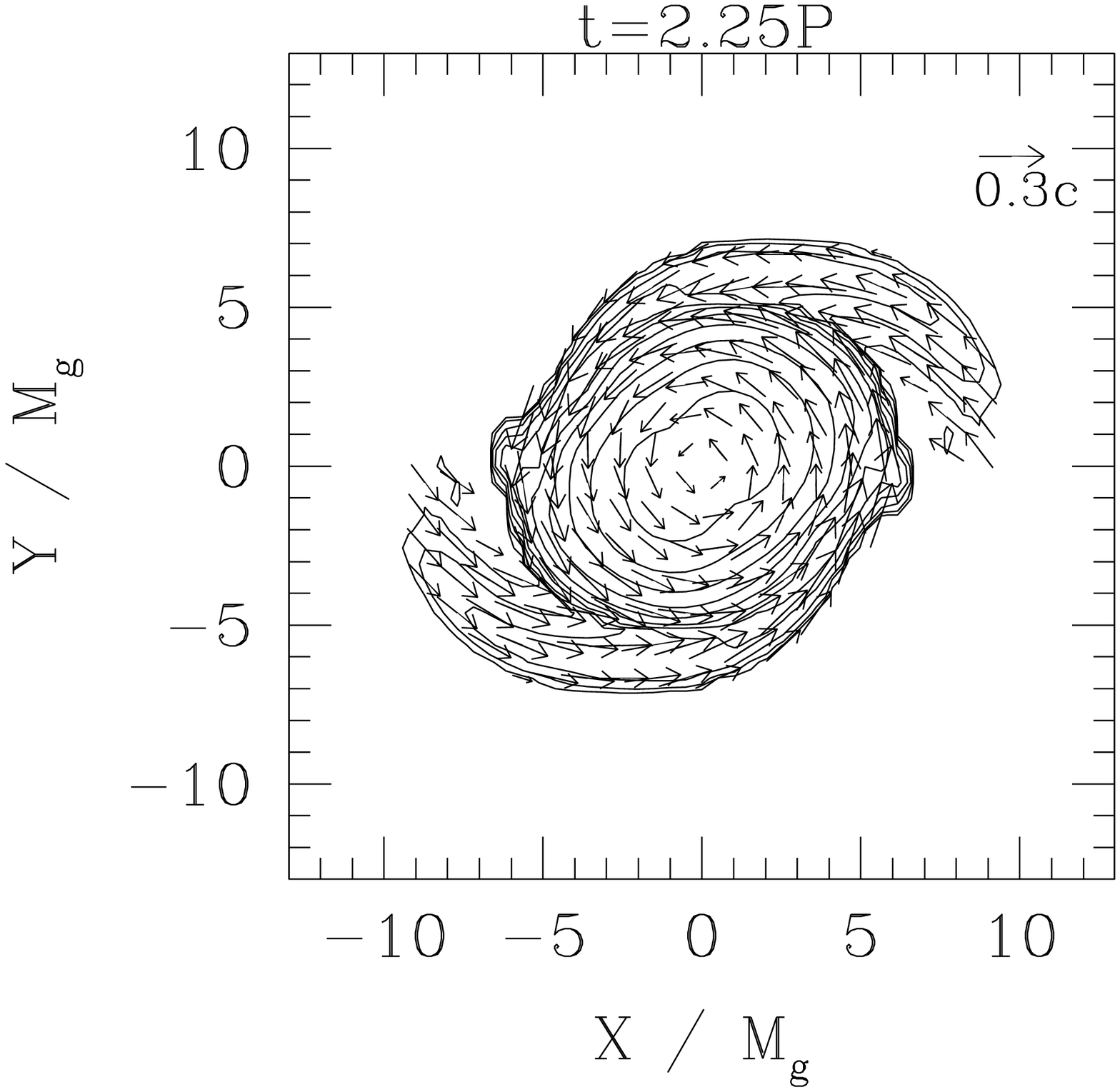}
\epsfxsize=1.8in
\leavevmode
\epsffile{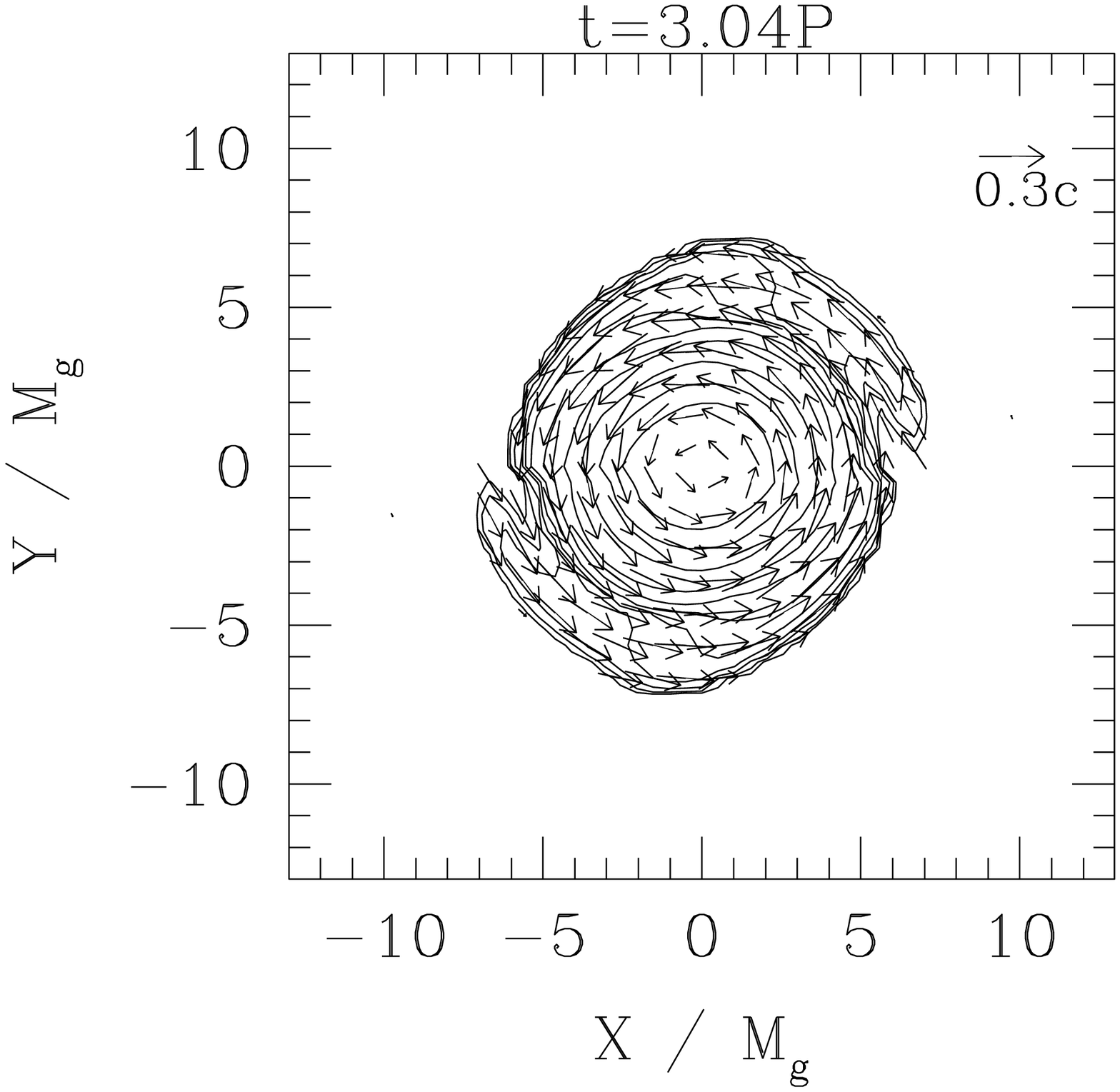} 
\caption{\footnotesize 
Snapshots of the density contours for $\rho_*$ 
in the equatorial plane for model (A). 
The contour lines are drawn for 
$\rho_*/\rho_{*~{\rm max}}=10^{-0.3j}$, 
where $\rho_{*~{\rm max}}$ denotes the maximum value of $\rho_*$ at 
$t=0$ (here $\bar \rho_{*~\rm max} = 0.355$), for $j=0,1,2,\cdots,10$. 
The maximum density for $\rho_*$ in the final panel is about 2.8 times 
larger than the initial value. 
Vectors indicate the local velocity field and the scale 
is as shown in the top left-hand frame. 
$P$ denotes the orbital period of the initial quasiequilibrium 
($P_{t=0}$). 
The length scale is shown in units of $GM_{g}/c^2$ 
where $M_g$ is the gravitational mass at $t=0$.\label{fig:FIG1} }
\end{center}
\vspace{-6mm}
\end{figure}

\begin{figure}[t]
\begin{center}
\epsfxsize=1.8in
\leavevmode
\epsffile{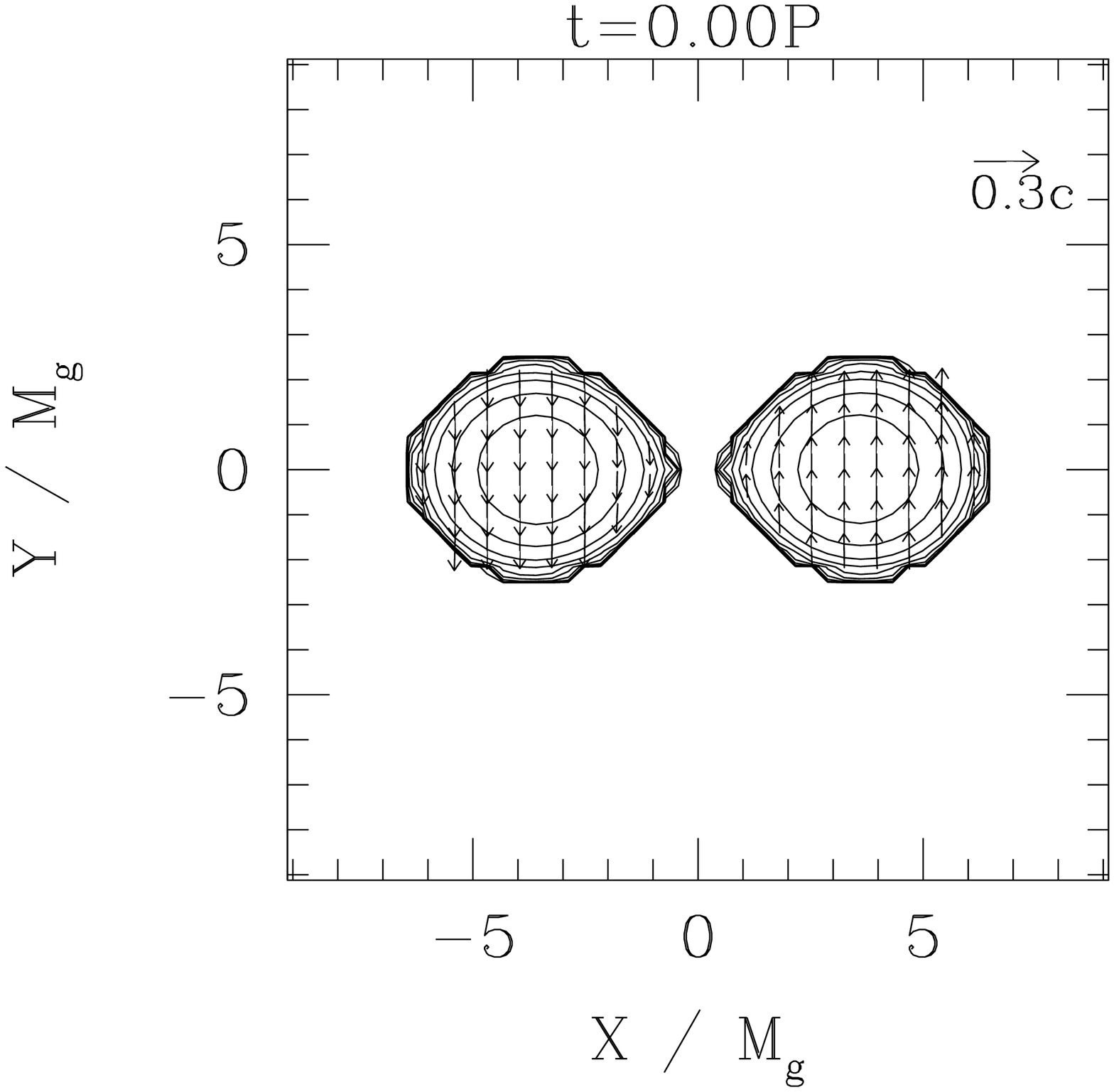}
\epsfxsize=1.8in
\leavevmode
\epsffile{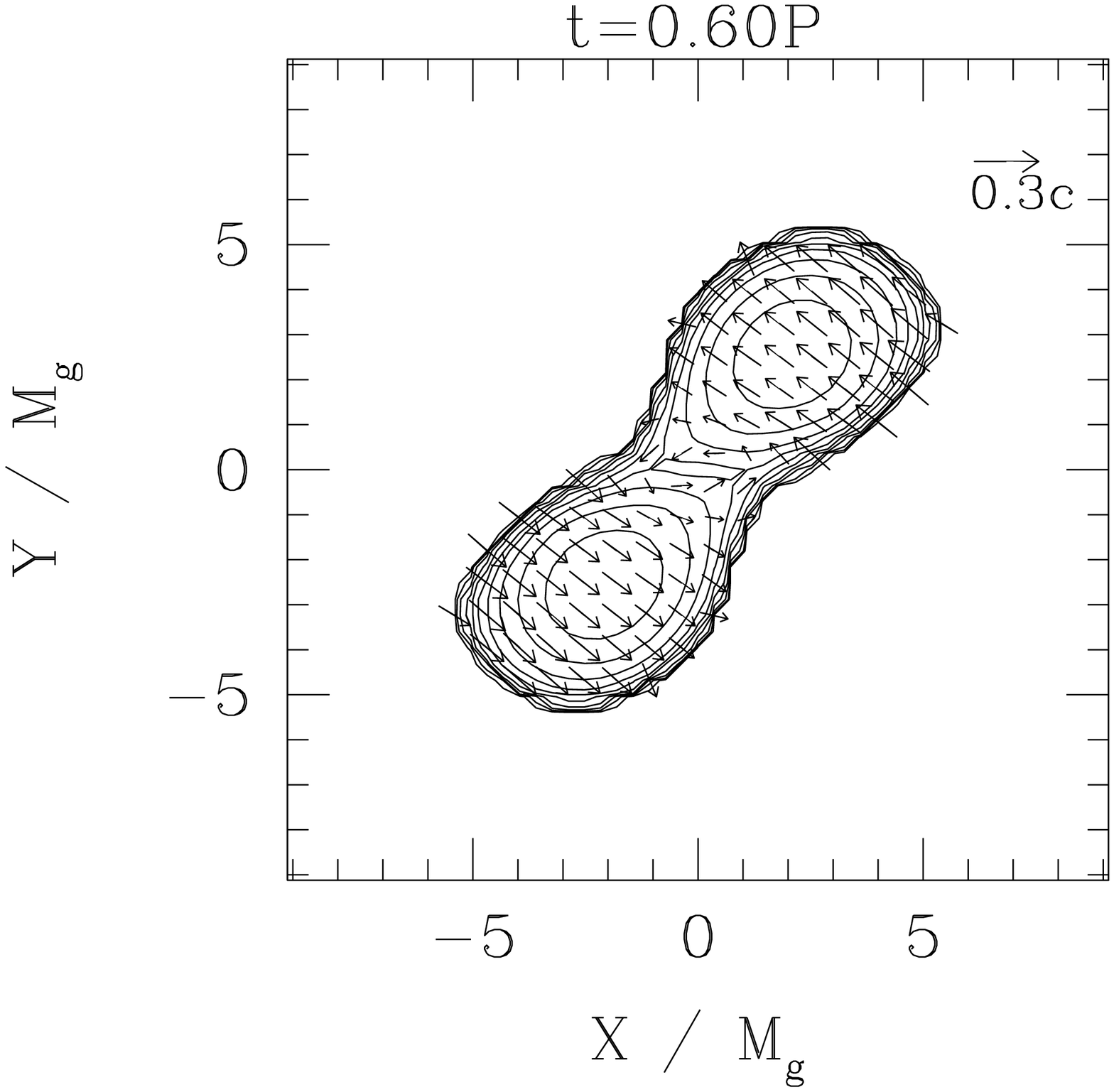} 
\epsfxsize=1.8in
\leavevmode
\epsffile{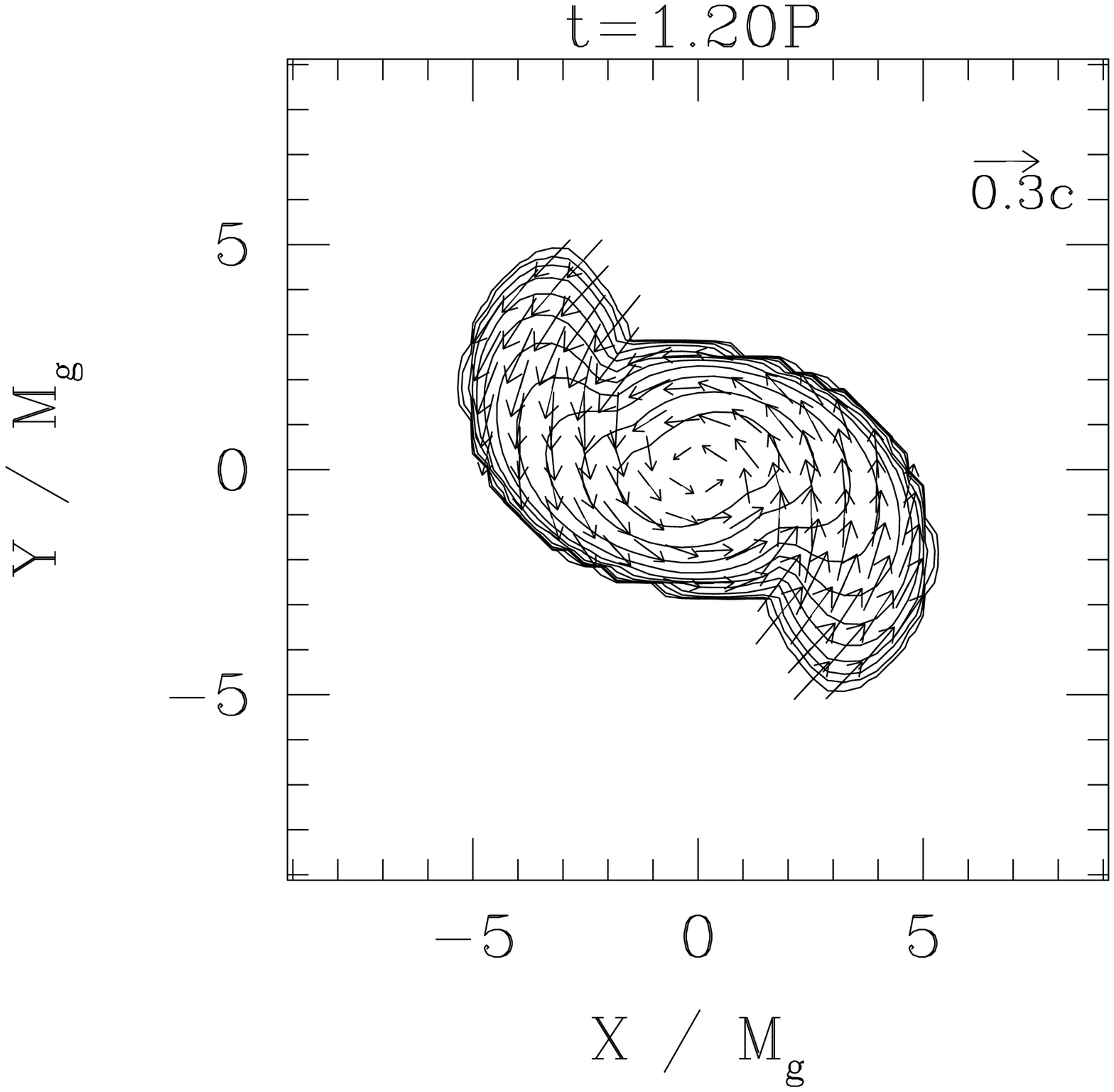} \\
\epsfxsize=1.8in
\leavevmode
\epsffile{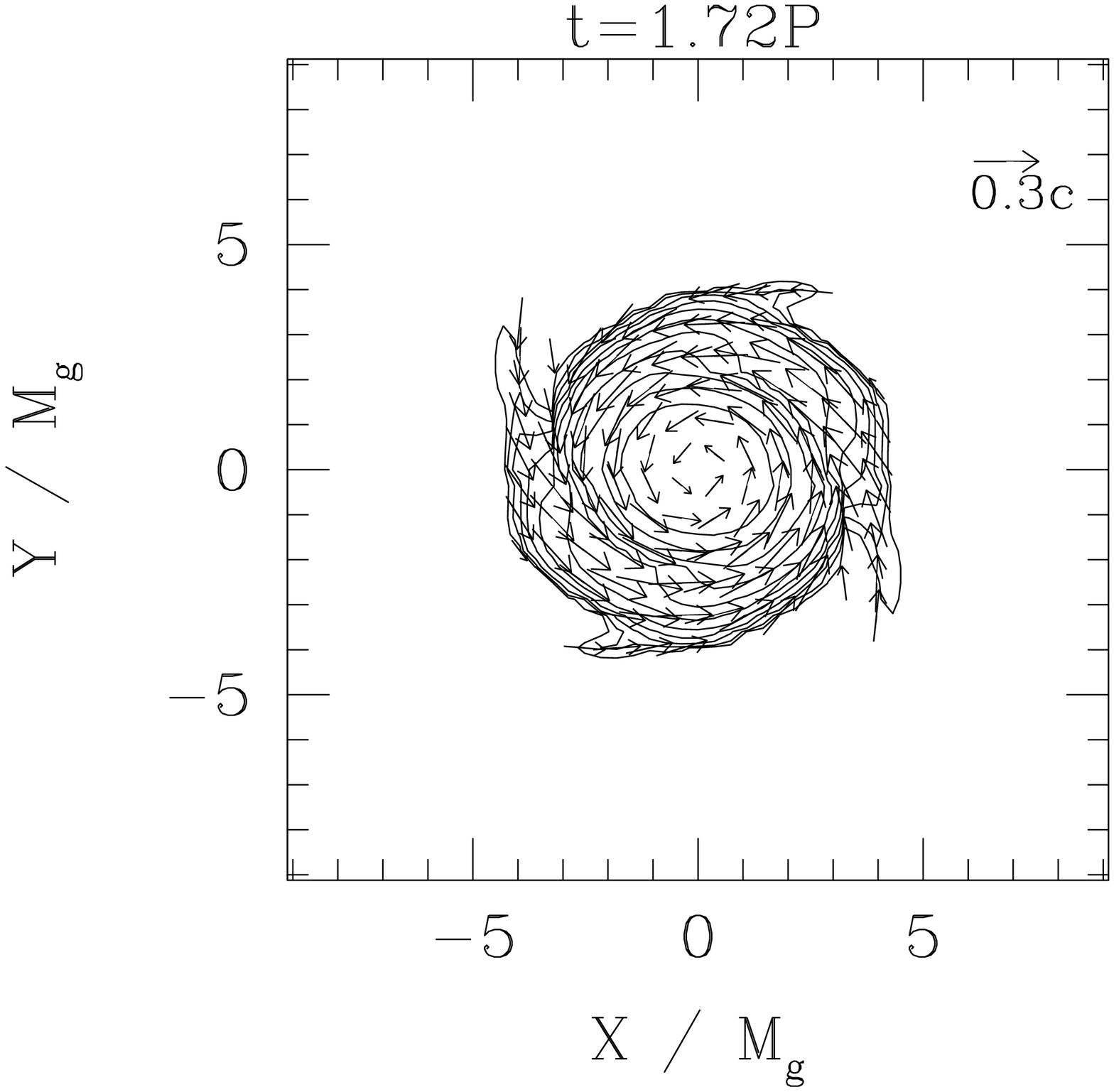} 
\epsfxsize=1.8in
\leavevmode
\epsffile{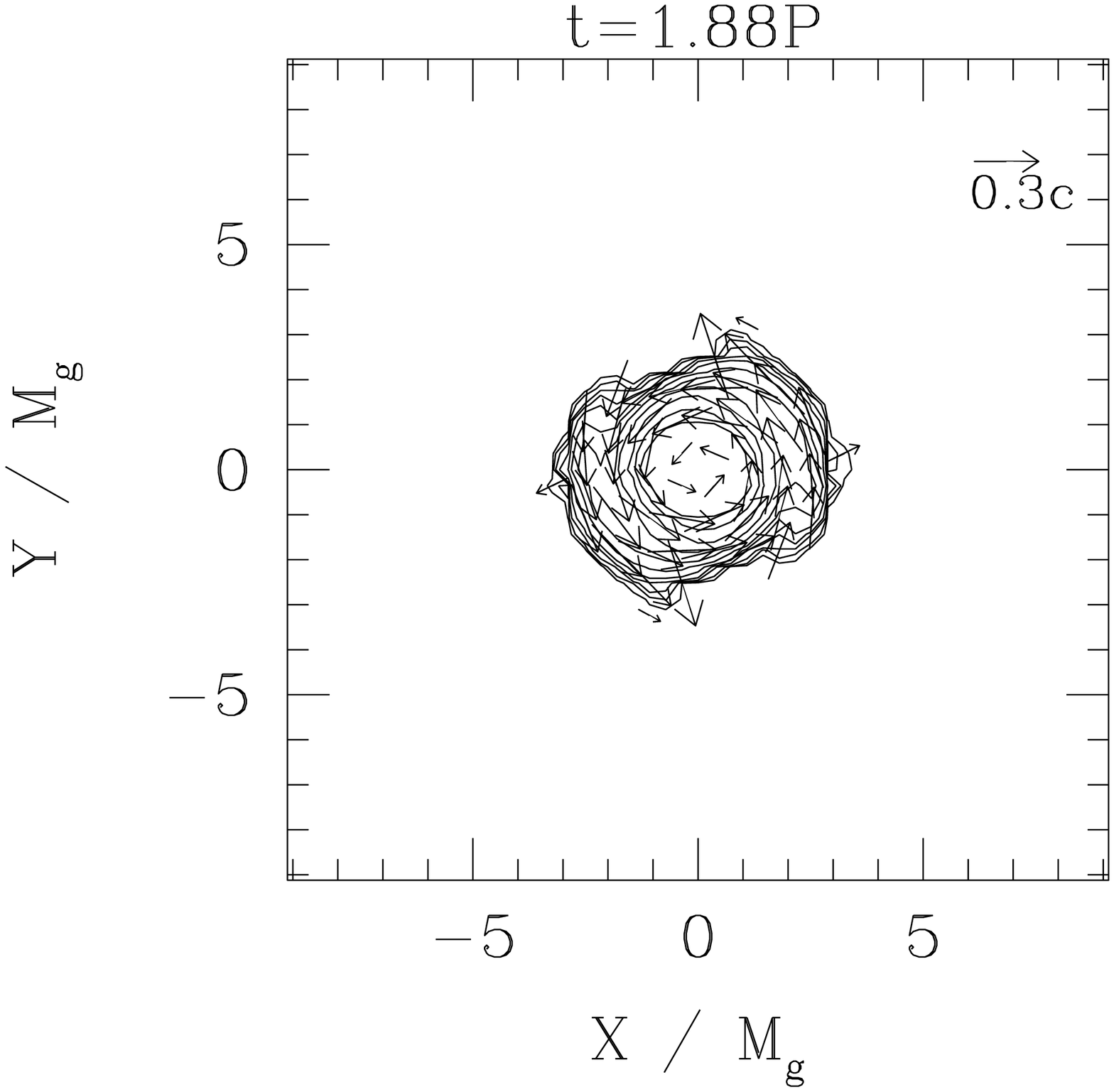}
\epsfxsize=1.8in
\leavevmode
\epsffile{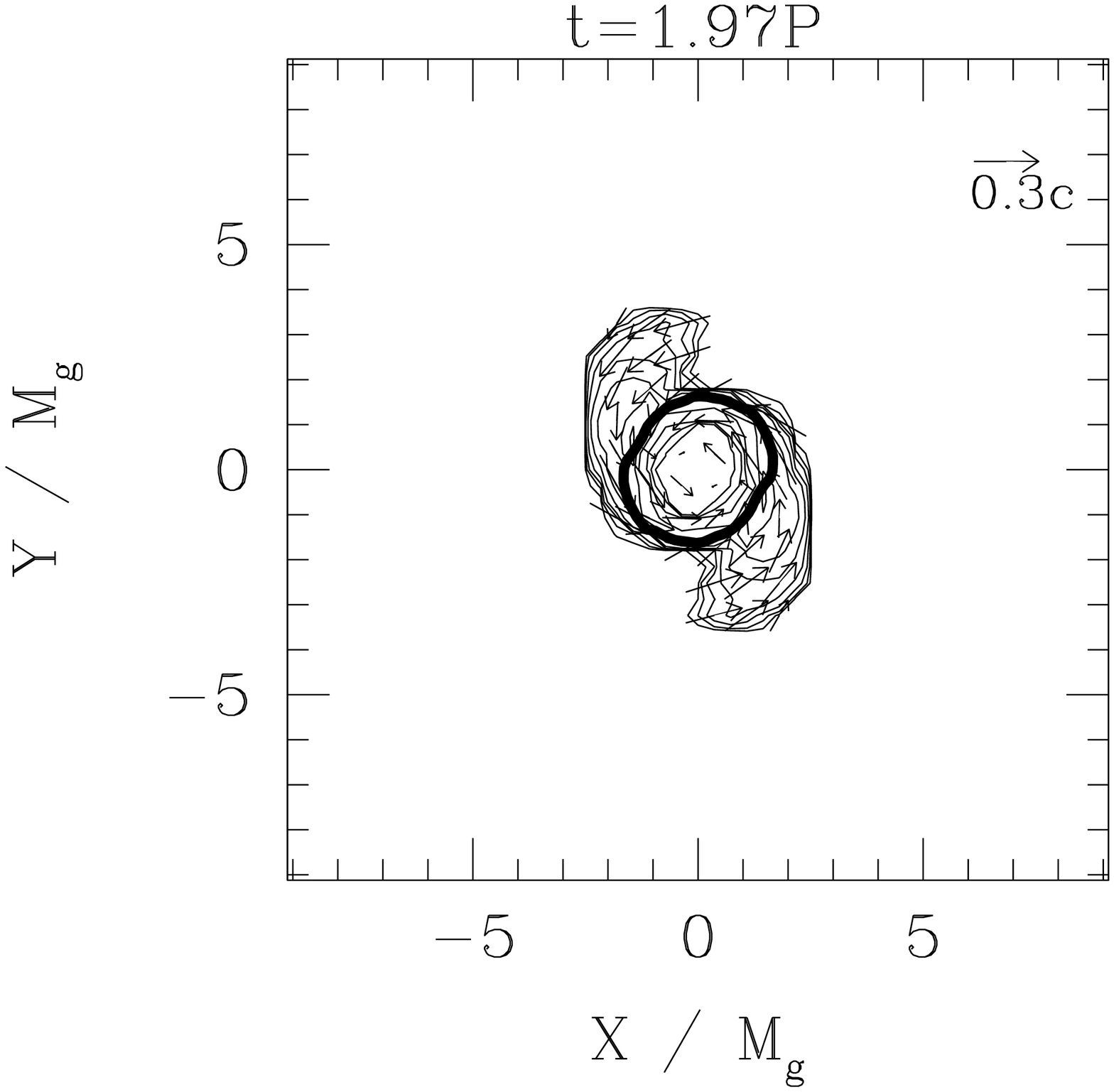} 
\caption{\footnotesize The same as \ref{fig:FIG1}, but 
for model (C). The contour lines are drawn for 
$\rho_*/\rho_{*~{\rm max}}=10^{-0.3j}$, 
where $\bar \rho_{*~{\rm max}}=0.757$, for $j=0,1,2,\cdots,10$. 
The maximum density for $\rho_*$ in the final panel is about 80 times 
larger than the initial value. 
The thick solid circle in the final panel denotes the apparent horizon. 
\label{fig:FIG3}}
\end{center}
\vspace{-6mm}
\end{figure}

In Figs. 1 and 2, we display the density contour lines 
and velocity vectors for $\rho_*$ and $v^i$ 
at selected timesteps for simulations of models (A) and (C). 
For (A) the final product is a massive neutron star. For (C) 
a black hole is formed and the apparent 
horizon is able to be located (thick solid circle in 
the last panel). 

\begin{figure}[htb]
\begin{center}
\epsfxsize=2.4in
\leavevmode
{\footnotesize (a)}\epsffile{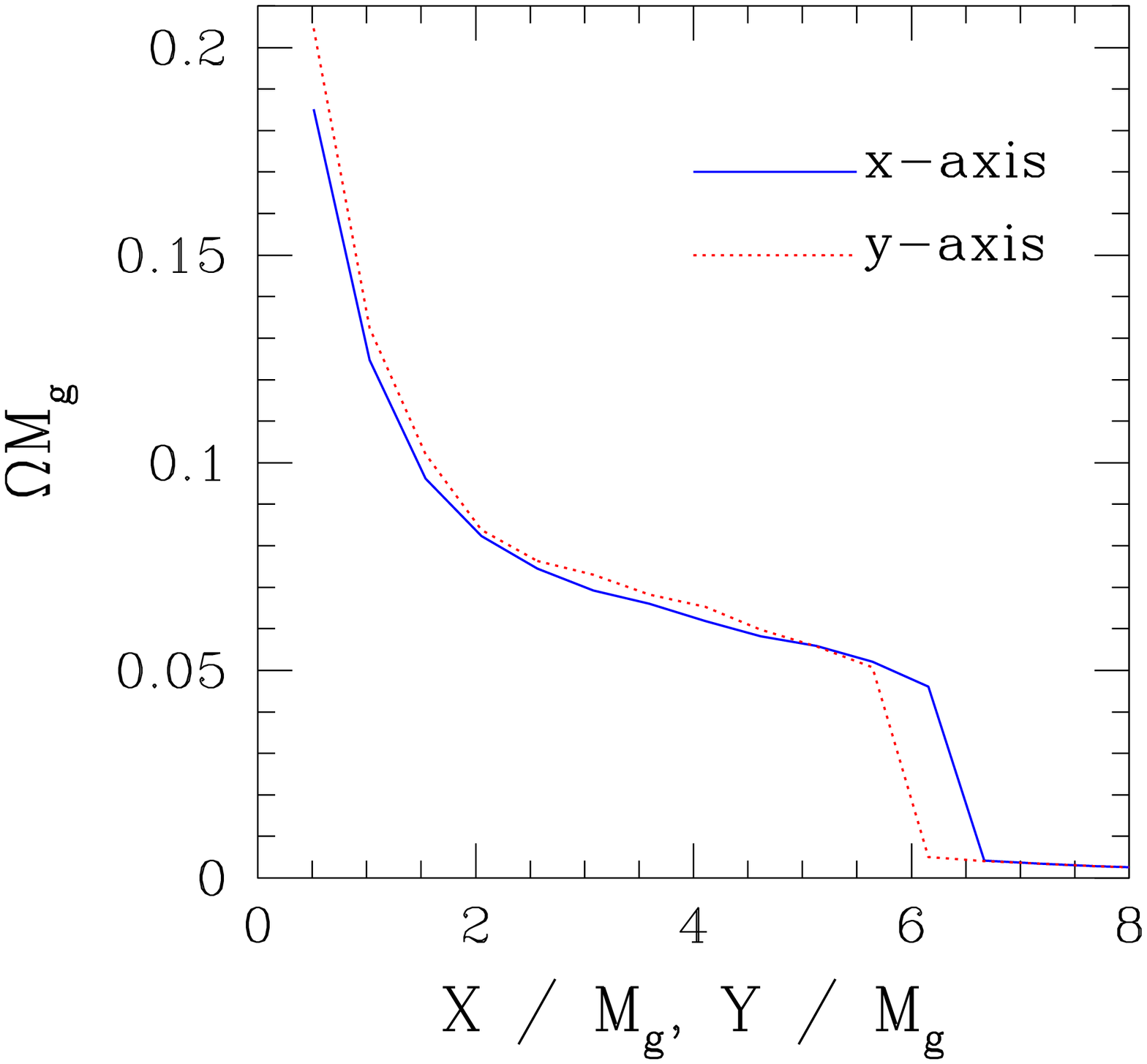}
\epsfxsize=2.4in
\leavevmode
~~~~~~~~{\footnotesize (b)}\epsffile{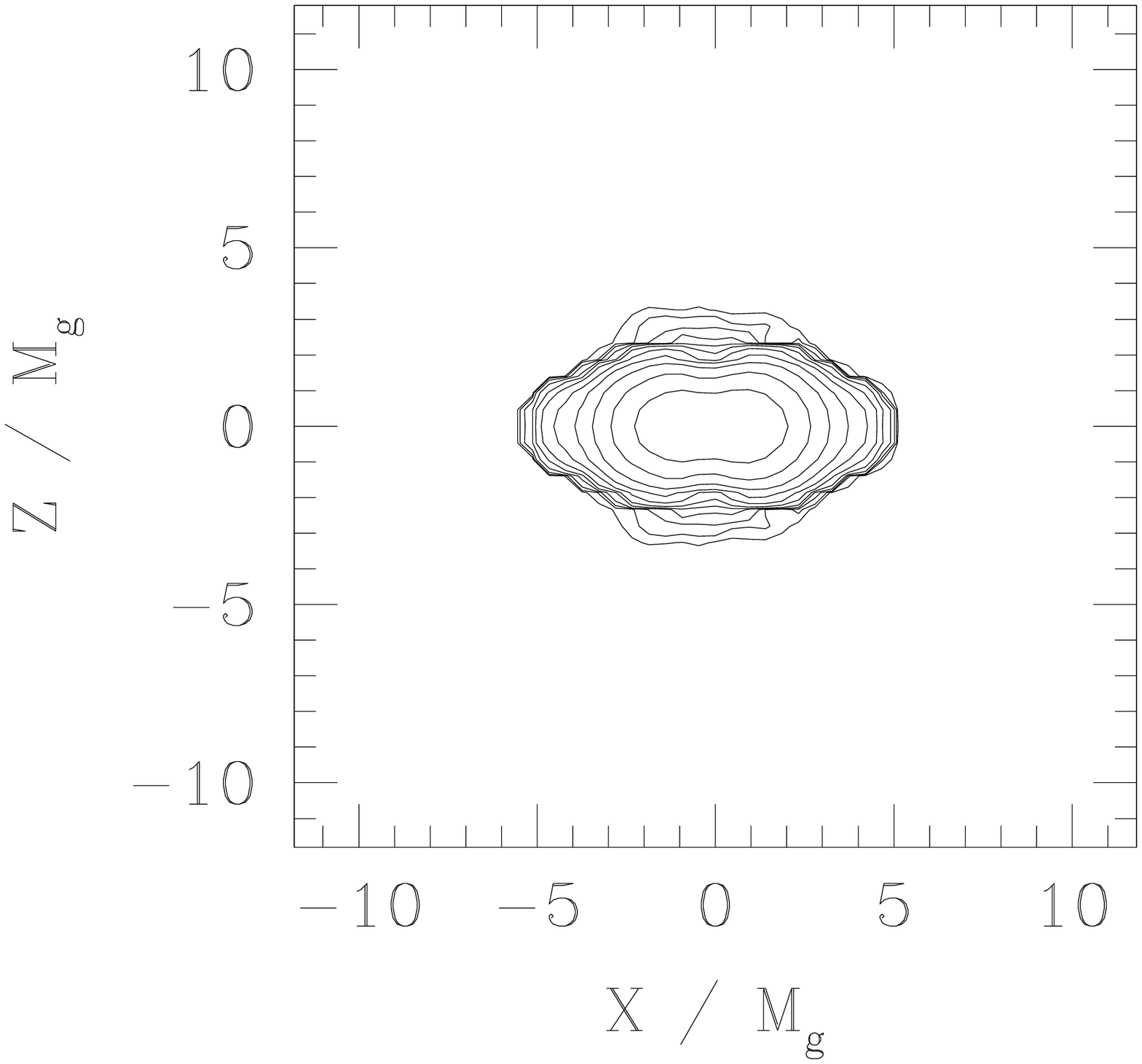}
\caption{\footnotesize (a) The angular velocity in units 
of $M_g^{-1}$ along $x$ and $y$ axes and (b) 
the density contour in $x$-$z$ slices 
at $t=3.04P_{t=0}$ for the merged object of model (A). 
The contour lines are drawn in the same manner as for Fig. 1. 
\label{fig:FIG4}}
\end{center}
\vspace{-6mm}
\end{figure}

\begin{figure}[htb]
%%\centerline{\hbox{ \psfig{figure=gw225.ps} }}
\begin{center}
\epsfxsize=2.4in
\leavevmode
{\footnotesize Fig. 4}\epsffile{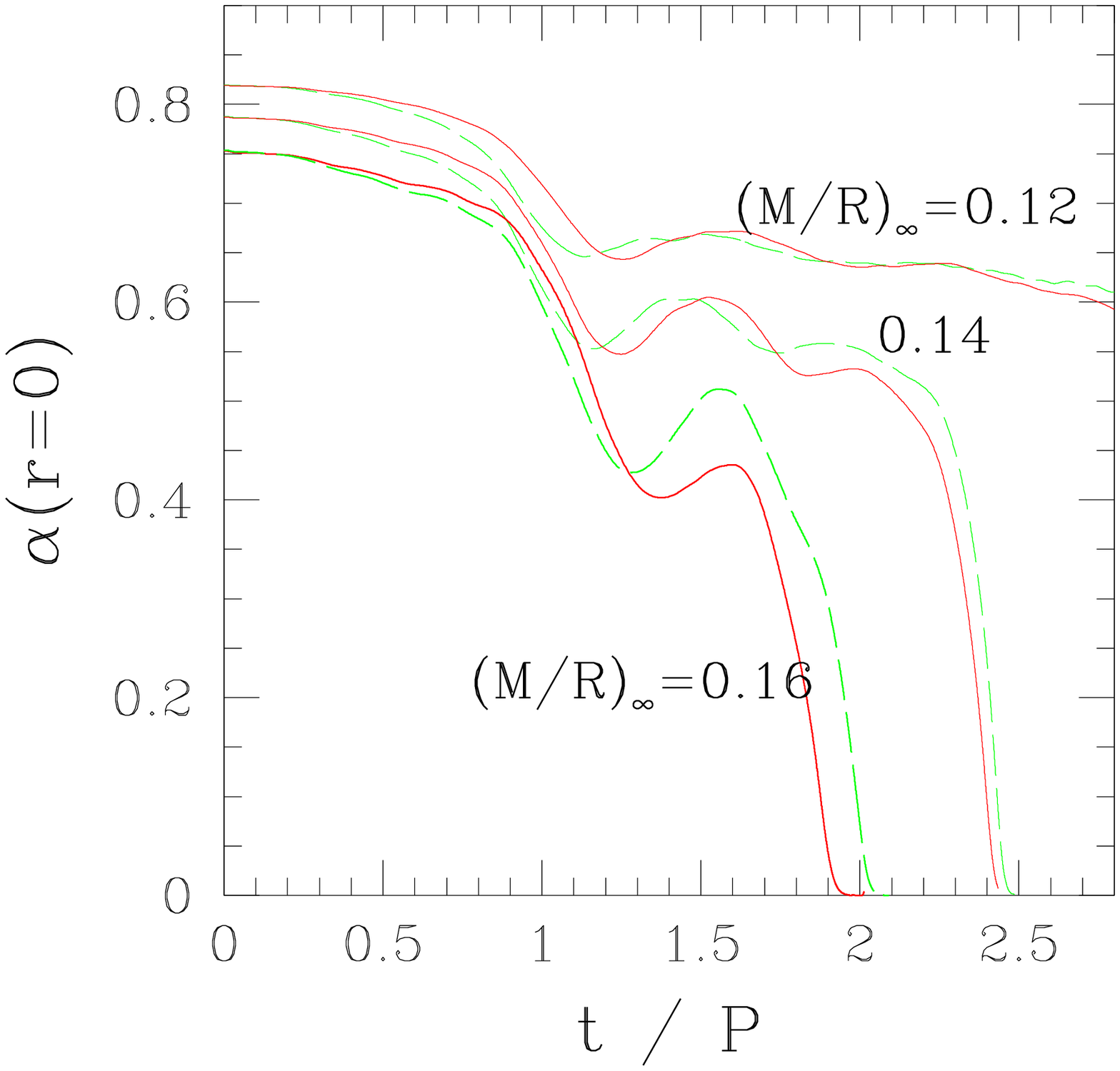}
\epsfxsize=2.4in
\leavevmode
{\footnotesize Fig. 5}\epsffile{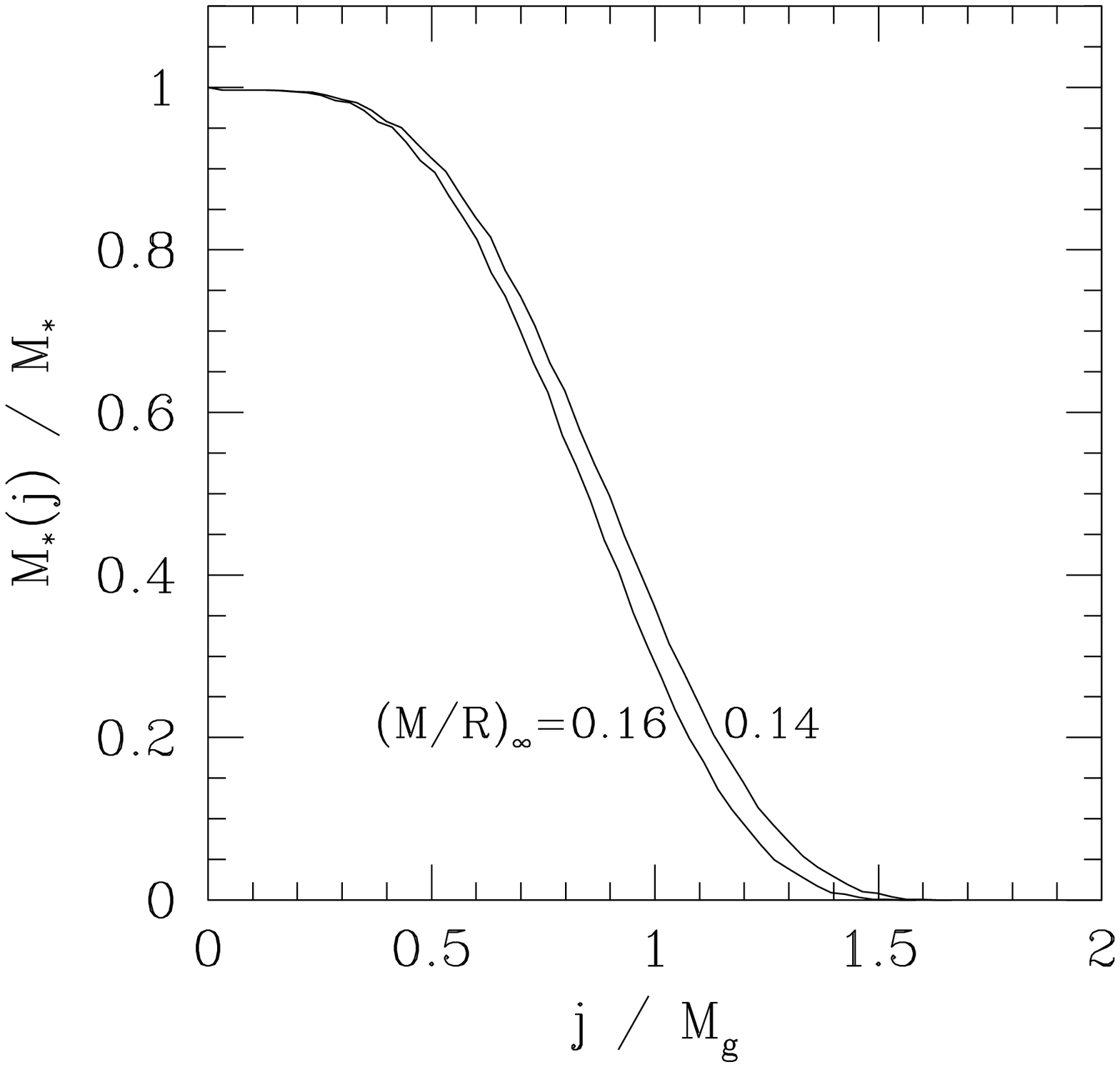}
\caption{\footnotesize The lapse function $\alpha$ at $r=0$ 
as a function of coordinate time for models (A)--(C). 
The solid and dashed lines denote the results for simulations with 
$(293,293,147)$ grid resolution and with 
$(193,193,97)$ grid resolution. 
For smaller scale simulations, the grid size is $116/96$ times 
larger than that for larger scale simulations. 
(Namely, the outer boundaries along each axis 
are located $116L/146$ in these cases.) 
\label{fig:FIG5}}
\caption{\footnotesize $M_*(j)/M_*$ as a function of $j/M_g$ 
for quasiequilibrium configurations (B) and (C). \label{fig:FIG6}}
\end{center}
\vspace{-6mm}
\end{figure}

\begin{figure}[htb]
\begin{center}
\epsfxsize=2.6in
\leavevmode
{\footnotesize (a)}\epsffile{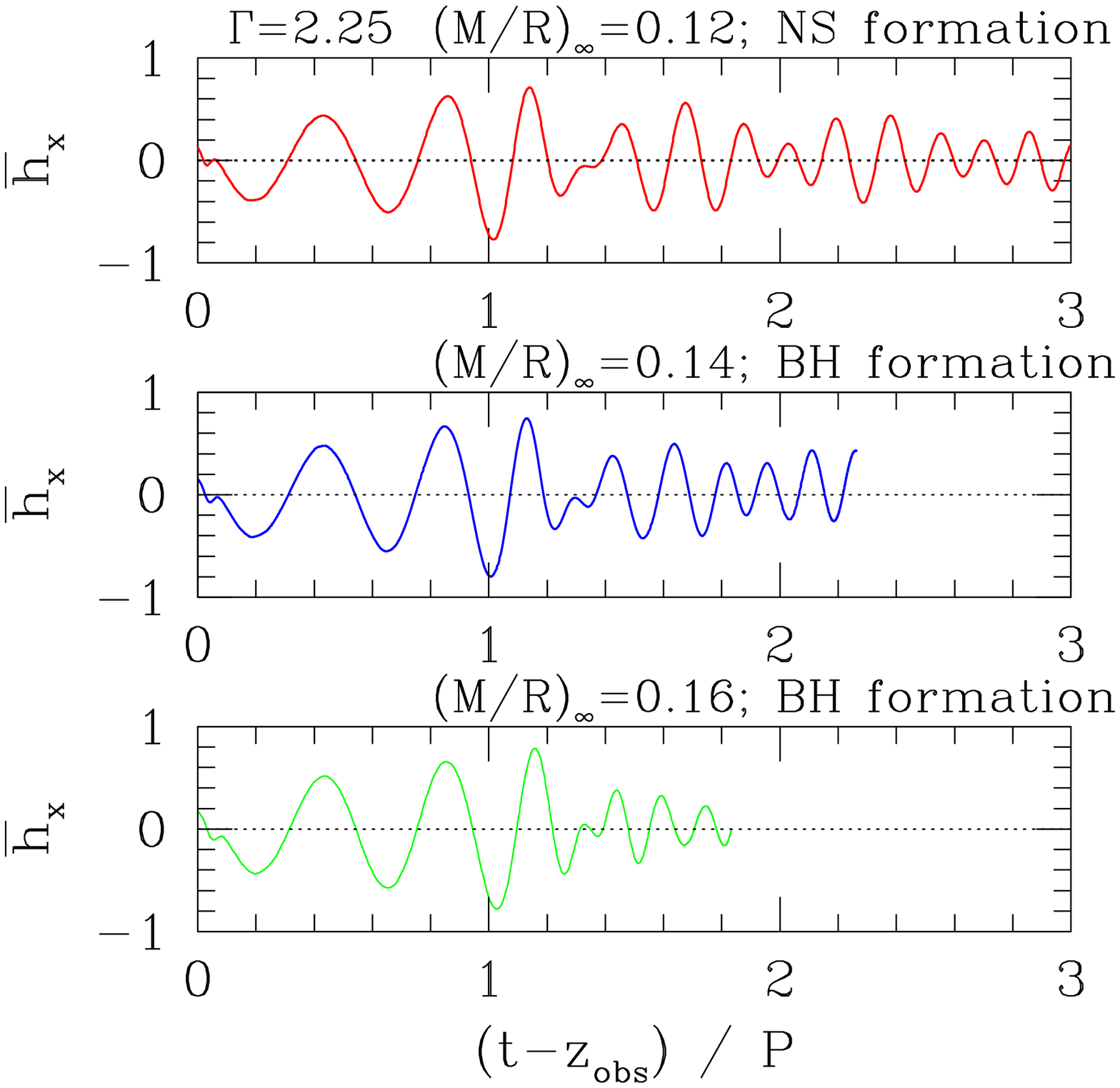}
\epsfxsize=2.5in
\leavevmode
~~~~{\footnotesize (b)} \epsffile{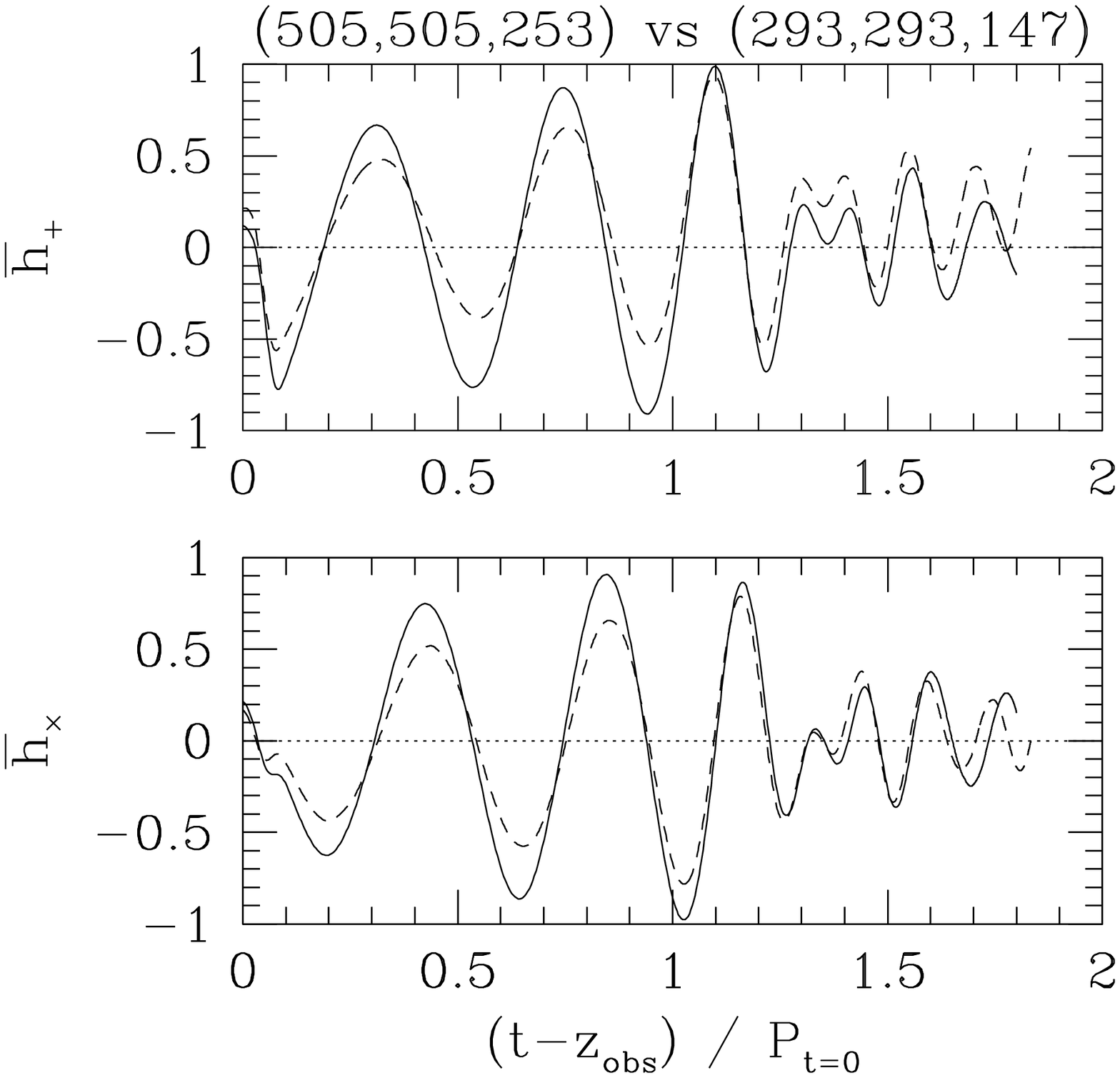}
\caption{\footnotesize (a) $\bar h_{\times}$ 
near the outer boundary along the $z$-axis 
as a function of retarded time (in units of $P$) for models (A)--(C) with 
$(293,293,147)$ grid size. $P$ denotes $P_{t=0}$. 
(b) $\bar h_{+}$ and $\bar h_{\times}$ for model (C) with 
$(505,505,253)$ (solid line) and 
$(293,293,147)$ (dashed line) grid sizes. }
\end{center}
\vspace{-6mm}
\end{figure}

The total rest mass of the binary for model (A) is $\sim 20\%$ larger than the 
maximum allowed value of a spherical star in isolation. 
Even with such large mass, the merged object does not collapse to
a black hole. 
As indicated in Fig. 1, the merger proceeds very mildly, 
because the approaching velocity at the contact of two stars is not 
very large. Consequently, 
the shock heating is not very important in merging. 
%%Indeed, the increase factor of $P/\rho^{\Gamma}$ from initial value 
%%(i.e., $\kappa$), which indicates the increase of the entropy, 
%%is quite small (at most 10$\%$) around the central region of the merged 
%%object. 
This implies that the rotational centrifugal force plays an 
important role for supporting the self-gravity of such supramassive 
outcome. To illustrate the importance of 
the rotation for the supramassive neutron star, the angular velocity along 
$x$ and $y$ axes and density contour lines in $x$-$z$ slices are displayed 
in Fig. 3. It is found that the supramassive neutron star is 
differentially rotating and consequently has a highly flattened configuration. 
Fig. 3(a) shows that the magnitude of the 
angular velocity in the inner region 
is of order of the Kepler velocity, i.e., $\Omega M_g=O[(M_g/R)^{3/2}]$ 
where $R$ denotes the typical radius of the merged object.

All these results are qualitatively the same as those 
found in simulations for $\Gamma=2$ \cite{bina}. 
However, the results are quantitatively different. 
For $\Gamma=2$, the maximum allowed rest mass for 
formation of a massive neutron star after merger is 
$\sim 1.5M_{*\rm max}^{\rm sph}$ where 
$M_{*\rm max}^{\rm sph}$ denotes the maximum allowed rest mass of 
a spherical star in isolation for a fixed value of $\Gamma$. 
As shown here, the threshold is slightly smaller as 
$\sim 1.3-1.4M_{*\rm max}^{\rm sph}$ for $\Gamma=2.25$ 
(we determine that model (B) is located near the threshold; see below), 
and we have found that it is $\sim 1.6-1.7M_{*\rm max}^{\rm sph}$
for $\Gamma=1.8$. Thus, the threshold value depends 
on the stiffness of the equation of state. 

It should be noted that the supramassive neutron star formed 
is temporarily stable, but not forever. It will eventually collapse 
to a black hole as argued in \cite{rot2}.

For models (B) and (C) in which $C \geq 0.14$, 
a black hole is formed. 
However, the formation process is 
slightly different between models (B) and (C). 
For model (C), a black hole is quickly formed after the first contact 
of two stars. On the other hand, for model (B), 
the merged object quasi-radially oscillates for a couple of times 
after the first contact instead of prompt formation of black hole. 
Indeed, as shown in Fig. 4, the lapse function at $r=0$ does not 
quickly approach to zero in this case. 
The collapse toward a black hole seems to occur 
after dissipating the angular momentum by 
gravitational radiation. The difference 
with regard to the formation process of black holes is reflected 
in gravitational waveforms (cf. Fig. 6). 

In Fig. 4, we also show time evolution of 
$\alpha$ at $r=0$ for simulations with (193,193,96) grid size.
It is found that the results do not contradict with those with 
(293,293,147) grid size, and 
convergence is achieved fairly well. However,  
with lower resolution, the numerical dissipation of 
the angular momentum is larger 
so that the merger happens earlier than that for higher resolution. 
Also, high density peaks are captured less accurately because 
of larger numerical diffusion. 
For even lower resolution, 
the merger for model (B) could avoid immediate collapse and form 
a massive neutron star because of large diffusion. 
Thus, we deduce that the total rest mass of model (B) is 
near a threshold for formation of black hole. 
To determine the threshold of the total rest mass for 
black hole formation more accurately, 
better-resolved numerical simulations are necessary. 

In the outcome of model (C), 
the rest mass outside the apparent horizon is less than $1\%$ 
of the total. This implies that the 
rest mass of the disk around the black hole is very small. 
For model (B), we were not able to determine the apparent 
horizon before the computation crashed. However, we found that 
the mass fraction 
outside spheres of a fixed coordinate radius (e.g., $r=1.5$ and $3M_g$) 
is decreasing with time to be very small. Thus, we expect that 
the mass of the disk is also very small in this case. 
In the following, we describe the reason for these results.

In Fig. 5, we show the mass spectrum with respect to the 
specific angular momentum 
$M_*(j)/M_*$ at $t=0$ \cite{USE} for models (B) and (C). 
Here, $j$ is the specific angular momentum 
$(1+\varepsilon+P/\rho)u_{\varphi}$ and $M_*(j)$ is defined as
\beq
M_*(j)=\int_{j'>j}d^3x' \rho_*(x')~~{\rm and}~~M_*(0)=M_*. 
\eeq
It is found that there is no fluid element for which $j/M_g > 1.6$. 
	
As we found in the simulations, quite a large fraction of the fluid elements 
are swallowed in the black hole. 
Gravitational radiation carries the energy from the system 
in particular in the early phase, 
but it should be less than $1\%$ of $M_g$ according to the quadrupole 
formula (see, e.g., \cite{gw3p2}). Thus, the 
mass of the black holes is approximately equal to the initial value. 
On the other hand, the angular momentum may be dissipated 
by gravitational waves by about $10\%$ of the initial value. 
These facts imply that $q=J/M_g^2$ should slightly decrease 
from the initial value to be $q \sim 0.9$ for both models. 
The specific angular momentum of a test particle in the 
innermost stable circular orbit around a Kerr black hole of mass 
$M_g$ and $q=0.9$ (0.95) is $\simeq 2.1M_g$ $(1.9M_g)$. 
Therefore, {\it any fluid element of irrotational binary 
neutron stars just before the merger does not have large specific 
angular momentum enough to form a disk around the formed black hole.}
For the disk formation, certain transport mechanism of the angular 
momentum such as hydrodynamic interaction is necessary. 
Since the black holes are formed in the dynamical timescale of 
the system, the mechanism has to be very effective 
to transport the angular momentum by more than $30\%$ in 
such short timescale. However, 
such rapid process is unlikely to happen as 
indicated in the present simulations. 

To observe gravitational waveforms, we extract $+$ and $\times$ 
gravitational waves along $z$-axis defined as 
\beq
\bar h_{+}\equiv {\tilde \gamma_{xx}-\tilde \gamma_{yy} \over 2}
\biggl({r \over M_g}\biggr)
\biggl({M \over R}\biggr)_{\infty}^{-1},~{\rm and}~
\bar h_{\times}\equiv \tilde \gamma_{xy} \biggl({r \over M_g}\biggr)
\biggl({M \over R}\biggr)_{\infty}^{-1}. 
\eeq
In Fig. 6 (a), we show $\bar h_{\times}$ 
as a function of retarded time near the outer boundary for models 
(A)--(C) with $(293,293,147)$ grid size. 
To illustrate the effect of the location of outer boundaries, 
we show $\bar h_+$ and $\bar h_{\times}$ 
for simulations with $(505,505,253)$ and (293,293,147) grid sizes 
for model (C) in Fig. 6(b). 
Note again that outer boundaries along 
each axis reside inside the wave zone 
in the early stage of the simulation. This effect results in 
the underestimation of the wave amplitude for 
$t-z_{\rm obs} \alt P_{t=0}$ where $P_{t=0}$ is the orbital period 
at $t=0$ as shown in  Fig. 6(b) (compare results in two different 
grid sizes). 
According to a second post Newtonian study \cite{BIWW}, 
the maximum wave amplitude of $\bar h_{+,\times}$ at $t \sim 0$ 
should be $\sim 0.7$. 
Thus, with (293, 293, 147) grid size, the wave amplitude is underestimated 
by a factor of 2. However, with larger grid size in which 
$L/\lambda_{t=0} \sim 2/3$, the factor of the underestimation is $\sim 10\%$, 
indicating that fairly accurate waveform could be calculated with 
slightly larger grid size with $L/\lambda_{t=0} \sim 1$. 

For $t - z_{\rm obs} \agt P_{t=0}$, on the other hand, 
$L$ is smaller than gravitational wavelength because 
the characteristic wavelength becomes short after the merger starts. 
Therefore, the waveforms in the late phase are considered to be fairly 
accurate. Indeed, the wave amplitudes for two simulations of 
different grid sizes agree well. 

In the case of massive neutron star formation, 
quasi-periodic gravitational waves of a fairly large amplitude, which 
are excited due to the non-axisymmetric oscillations of the 
merged object, are emitted after the merger. 
Since the radiation reaction timescale is much longer than 
the dynamical (rotational) timescale of the system, 
the quasi-periodic waves will be emitted for many rotational cycles. 

Even for the case of black hole formation, 
quasi-periodic gravitational waves are excited 
due to the non-axisymmetric oscillation of merged objects before 
collapse to a black hole. 
Since the computation crashed soon after the formation of 
the apparent horizon, we cannot draw definite conclusion 
with regard to gravitational waves in the last phase. 
However, we can expect that after the formation of 
a black hole, its quasi-normal modes are  
excited and gravitational waves will damp eventually. 
Since the formation timescale of the black hole is different 
between models (B) and (C) depending on initial 
compactness of neutron stars, duration of the 
quasi-periodic waves induced by non-axisymmetric oscillations of 
the transient merged objects is also different. 
%%%From these results, the spectrum of gravitational 
%%%waves after the merging phase may be illustrated as in Fig. 6 (b). 
The wavelength of the quasi-periodic oscillation is $\sim 3$ 
times shorter than $\lambda_{t=0}$, and thus, the typical frequency 
can be estimated as 
\beq
f \simeq 2.0{\rm kHz}\biggl({2.8M_{\odot} \over M_g}\biggr)
\biggl({X \over 0.1}\biggr)^{3/2}. 
\eeq
The amplitude of the quasi-periodic oscillation in the Fourier domain 
is determined by its duration and hence, 
depends strongly on the initial compactness of neutron stars and 
final product. Therefore, by observing the amplitude of this peak, 
we will be able to obtain information about the compactness 
of neutron stars before the merger, and final product. 
From gravitational waves emitted in the inspiraling phase with 
post Newtonian templates of waveforms \cite{BIWW}, 
mass of two neutron stars, and hence, total mass 
will be determined \cite{CF}. This implies that 
we could constrain the maximum allowed 
mass of neutron stars, and hence, nuclear equations of state 
from the amplitude of the quasi-periodic oscillation emitted 
by the merged object. 

Since the frequency of this Fourier peak is rather high, it will be 
difficult to detect by first generation, kilo-meter-size 
laser interferometers such as LIGO. 
However, the resonant-mass detectors and/or 
specially designed narrow band interferometers may be available 
in future, to detect such high frequency gravitational waves. 
These detectors will provide us a variety of 
information on neutron star physics.

\section{Summary}

We have performed fully GR simulations of merger of binary 
neutron stars. As demonstrated in this paper, 
the simulations are feasible stably and fairly accurately 
to yield scientific results. 

One of the most interesting results found in this work is that 
the products after merger depend sensitively on the 
compactness of neutron stars before merger. If the total rest mass 
of the system is sufficiently (1.3--1.7 times depending on $\Gamma$) 
larger than the maximum rest mass of a spherical star in isolation, 
a black hole is formed, and otherwise, 
a massive neutron star is formed. It is noteworthy that 
the rest mass of the massive neutron star can be significantly larger than 
the maximum value for a spherical star of identical equation of 
state. The self-gravity of such high mass neutron stars can 
be supported by a rapid, differential rotation \cite{bina,rot2}. 
We also found that  
the difference of the final products is significantly 
reflected in the waveforms of gravitational waves, 
suggesting that detection of gravitational waves of high 
frequency could constrain the maximum allowed mass of neutron stars. 

In the case of prompt black hole formation, the disk mass is 
found to be very small, i.e., less than 1$\%$ of the total rest mass. 
The main reason is that the specific angular momentum of 
all the fluid elements in binary neutron stars 
of irrotational velocity field just before the merger is 
too small and transport timescale of the 
angular momentum is not short enough to help the disk formation. 
It should be noted that 
this conclusion may hold only for binary neutron stars of 
equal (or nearly equal) mass. 
In binaries of a large mass ratio, 
the conclusion could be modified, because 
the neutron star of smaller mass may be 
tidally disrupted before the separation of two stars becomes small 
and hence, before the angular momentum of the system is not 
significantly dissipated 
by gravitational radiation. In this case, many of fluid elements in  
the neutron star of smaller mass may have a large angular momentum 
enough to form a disk during the merger. To clarify whether such scenario is 
promising or not, it is necessary to perform simulations 
for merger of binary neutron stars of unequal mass. 

\vspace{-2mm}
\section*{Acknowledgments}

We would like to thank J. C. Wheeler 
for inviting M.S. to this meeting to give an opportunity for presenting 
our works. We also thank T. Baumgarte, E. Gourgoulhon, 
T. Nakamura, K. Oohara and S. Shapiro for discussions. 
Numerical computations were performed on the 
FACOM VPP300/16R, VX/4R and VPP5000 machines 
in the data processing center of National Astronomical Observatory 
of Japan. 

%%%%%%%%%%%%%%%%%%%%%%%%%%%%%%%%%%%%%%%%%%%%%%%%%%%%%%%%%%%%%%%%%%%%%%%%%%%
%%%           References starts here                                      %
%%%%%%%%%%%%%%%%%%%%%%%%%%%%%%%%%%%%%%%%%%%%%%%%%%%%%%%%%%%%%%%%%%%%%%%%%%%
\vspace{-4mm}
%\newpage
%\addcontentsline{toc}{section}{References}
%\begin{thebibliography}{99}

%\end{thebibliography}
\end{document}